\documentclass{article}
\usepackage{graphicx}
\usepackage{amsmath}
\usepackage{amsfonts}
\usepackage{amssymb}

\begin{document}

\title{\textbf{Classical Solutions in a} \textbf{Lorentz-violating
Maxwell-Chern-Simons Electrodynamics}}
\author{H. Belich Jr.$^{a,b}$, M.M. Ferreira Jr.$^{b,c}$, J.A. Helay\"{e}l-Neto$^{a,b}%
$ and M.T.D. Orlando$^{b,d}$\thanks{e-mails: belich@cbpf.br, manojr@cbpf.br,
helayel@cbpf.br, orlando@cce.ufes.br}\\$^{a}$\textit{Centro Brasileiro de Pesquisas F\'{i}sicas (CBPF)},\\Coordena\c{c}\~{a}o de Teoria de Campos e Part\'{i}culas (CCP), \\Rua Dr. Xavier Sigaud, 150 - Rio de Janeiro - RJ 22290-180 - Brazil.\\$^{b}$\textit{Grupo de F\'{i}sica Te\'{o}rica Jos\'{e} Leite Lopes, }\\Petr\'{o}polis - RJ - Brazil.\\$^{c}$\textit{Universidade Federal do Maranh\~{a}o (UFMA)}, \\Departamento de F\'{i}sica, Campus Universit\'{a}rio do Bacanga,\\S\~{a}o Luiz - MA, 65085-580 - Brazil. \\$^{d}$ \textit{Universidade Federal do Esp\'{i}rito Santo (UFES)},\\Departamento de F\'{i}sica e Qu\'{i}mica, Av. Fernando Ferrarim, S/N\\Goiabeiras, Vit\'{o}ria - ES, 29060-900 - Brasil}
\maketitle
\begin{abstract}
We take as starting point the planar model arising from the dimensional
reduction of the Maxwell Electrodynamics with the (Lorentz-violating)
Carroll-Field-Jackiw term. We then write and study the extended Maxwell
equations and the corresponding wave equations for the potentials. The
solution to these equations show some interesting deviations from the usual
MCS Electrodynamics, with background-dependent correction terms. In the case
of a time-like background, the correction terms dominate over the MCS
sector\ in the region far from the origin, and establish the behaviour of a
massless Electrodynamics (in the electric sector). In the space-like case, the
solutions indicate the clear manifestation of spatial anisotropy, which is
consistent with the existence of a privileged direction is space. \ 
\end{abstract}

\section{\ Introduction\-}

The intensive development of Lorentz- and CPT-violating theories in $\left(
1+3\right)  $-dimensions \cite{Jackiw}- \cite{Andrianov} has come across the
question about the structure of a similar model in 1+2 dimensions and its
possible implications. A dimensional reduction (to $D=1+2)$ of a
Lorentz-violating Maxwell Electrodynamics, based on presence of the
Carroll-Field-Jackiw term ($\epsilon^{\mu\nu\kappa\lambda}v_{\mu}A_{\nu
}F_{\kappa\lambda}$), has been recently accomplished \cite{Manojr}, resulting
in a gauge invariant Planar Quantum Electrodynamics (QED$_{3}$) composed by a
Maxwell-Chern-Simons gauge field $\left(  A_{\mu}\right)  ,$ by a Klein-Gordon
massless scalar field $\left(  \varphi\right)  $, and a fixed 3-vector
$\left(  v^{\mu}\right)  $. The MCS-Electrodynamics is supplemented by a
mixing, Lorentz-violating term, consisting of the gauge field and the external
background, $v^{\mu}$. In this way, one has derived a Lorentz- and
CTP-violating planar theory, whose structure stems from a known counterpart
previously defined in 1+3 dimensions. Some of its general features, concerning
the physical consistency of this model were investigated. One has then
verified that the overall model is endowed with stability, while the causality
is spoiled by some modes. The unitarity, however, is preserved even at modes
that violate causality (whenever one adopts a purely space-like background),
in such a way this model may suffer a consistent quantization. This planar
model, in spite of violating Lorentz covariance in the particle frame
\cite{Colladay}, may conserve the CTP\ symmetry for a specific choice of
$v^{\mu}$ \cite{Manojr}.

In this letter, one focuses\ attention on the issue of the classical equations
of motion (the extended Maxwell equations) and wave equations (for the
potential $A^{\mu})$ derived from the reduced Lagrangian. The purpose here is
to investigate the effects of the Lorentz-violating background on the field
strengths and potentials generated in our planar QED$_{3}$. Initially, one
verifies that these equations have a similar structure to the usual MCS case,
supplemented by terms that depend on the background vector. Solving these
equations, we obtain solutions that differ from the MCS ones also by $v^{\mu}%
$-dependent correction-terms both for time- and space-like $v^{\mu}$. In the
time-like case, qualitative physical changes appear when one investigates the
asymptotic character of the solutions. The background seems to annihilate the
screening characteristic of a massive Electrodynamics, leading to a behaviour
typical of massless QED$_{3}$ (at least in the electric sector). Near the
origin, no qualitative modification takes place. In this case, no signal of
\ spatial anisotropy appears. On the other hand, adopting a space-like
$v^{\mu}$, the spatial-anisotropy becomes a manifest property of the
solutions. Induced by the external background, the anisotropy arises in the
form of terms (with clear dependence on the angle relative to the fixed
direction determined by the background, $\overrightarrow{v})$ that correct the MCS\ behaviour.

In short, this letter is outlined as follows. In Section II, we present the
basic features of the reduced model, previously developed in ref.
\cite{Manojr}. In Section III, the equations of motion, from which one derives
the wave equations for potentials and field strengths is displayed. In Section
IV, we solve the equations (in the static limit)\ for the time- and space-like
cases and discuss the results. In Section V, we conclude by presenting our
Final Remarks.

\section{The Lorentz-violating Planar Model}

Our planar Lorentz-violating model is attained by means a dimensional
reduction of a the Maxwell Lagrangian\footnote{Here one has adopted the
following metric conventions: $g_{\mu\nu}=(+,-,-,-)$ in $D=1+3,$ and
$g_{\mu\nu}=(+,-,-)$ in $D=1+2$. The greek letters (with hat) $\hat{\mu},$ run
from 0 to 3, while the pure greek letters, $\mu,$ run from 0 to 2.} (written
in 1+3 dimensions) supplemented by the Carrol-Field-Jackiw term\cite{Jackiw}:
\begin{equation}
\mathcal{L}_{1+3}=\biggl\{-\frac{1}{4}F_{\hat{\mu}\hat{\nu}}F^{\hat{\mu}%
\hat{\nu}}+\frac{1}{2}\epsilon^{\hat{\mu}\hat{\nu}\hat{\kappa}\hat{\lambda}%
}v_{\hat{\mu}}A_{\hat{\nu}}F_{\hat{\kappa}\hat{\lambda}}+A_{\hat{\nu}}%
J^{\hat{\nu}}\biggr\}, \label{action1}%
\end{equation}
where $v^{\mu}$ represents the external background and $A_{\hat{\nu}}%
J^{\hat{\nu}}$ is a additional term considering the coupling between the gauge
field and an external current. This model (in its free version) is gauge
invariant but does not preserve Lorentz and CPT symmetries relative to the
particle frame \cite{Jackiw}, \cite{Kostelec1}, \cite{Manojr}. Applying the
prescription of the dimensional reduction, described in ref. \cite{Manojr}, on
the eq. (\ref{action1}), one obtains the reduced Lagrangian: \ \ \ \ \ \
\begin{equation}
\mathcal{L}_{1+2}=-\frac{1}{4}F_{\mu\nu}F^{\mu\nu}+\frac{1}{2}\partial_{\mu
}\varphi\partial^{\mu}\varphi-\frac{s}{2}\epsilon_{\mu\nu k}A^{\mu}%
\partial^{\nu}A^{k}+\varphi\epsilon_{\mu\nu k}v^{\mu}\partial^{\nu}A^{k}%
-\frac{1}{2\alpha}\left(  \partial_{\mu}A^{\mu}\right)  ^{2}+A_{\mu}J^{\mu
}+\varphi J, \label{Lagrange2}%
\end{equation}
where the gauge-fixing term was added up after the dimensional reduction. The
scalar field, $\varphi,$ is a Klein-Gordon massless field and also acts as the
coupling constant that links the fixed $v^{\mu}$ to the gauge sector of the
model, by means of the new mixing-term:\ $\varphi\epsilon_{\mu\nu k}v^{\mu
}\partial^{\nu}A^{k}.$ In spite of being covariant in form, this term breaks
the Lorentz symmetry in the particle-frame (since the 3-vector $v^{\mu}$ is
not sensitive to particle Lorentz boost), behaving like a set of three
scalars. This reduced model does not necessarily jeopardize the CPT
conservation, which depends truly on the character of the fixed vector
$v^{\mu}$: there will occur conservation if one works with a purely space-like
external vector ($v^{\mu}=(0,\overrightarrow{v})$) , or breakdown, if $v^{\mu
}$ is purely time-like or otherwise \cite{Manojr}. Here, these results were
established under the assumption $\varphi$ is a scalar field\footnote{As
discussed in ref. \cite{Manojr}, if this field behaves like a pseudo-scalar,
the CPT conversation will be assured for a purely time-like $v^{\mu}$.}.

To evaluate the propagators related to Lagrangian (\ref{Lagrange2}), one
defines some new operators that form a closed algebra:
\[
Q_{\mu\nu}=v_{\mu}T_{\nu},\ \Lambda_{\mu\nu}=v_{\mu}v_{\nu},\ \ \Sigma_{\mu
\nu}=v_{\mu}\partial_{\nu},\ \Phi_{\mu\nu}=T_{\mu}\partial_{\nu},
\]
Lengthy algebraic manipulations yield the propagators as listed below:\
\begin{align}
\text{ }\langle A^{\mu}\left(  k\right)  A^{\nu}\left(  k\right)
\rangle\text{ }  &  =i\biggl\{-\frac{1}{k^{2}-s^{2}}\theta^{\mu\nu}%
-\frac{\alpha(k^{2}-s^{2})\boxtimes(k)+s^{2}\left(  v_{\alpha}k^{\alpha
}\right)  ^{2}}{k^{2}(k^{2}-s^{2})\boxtimes(k)}\omega^{\mu\nu}-\frac{s}%
{k^{2}(k^{2}-s^{2})}S^{\mu\nu}\nonumber\\
&  +\frac{s^{2}}{(k^{2}-s^{2})\boxtimes(k)}\Lambda^{\mu\nu}-\frac{1}%
{(k^{2}-s^{2})\boxtimes(k)}T^{\mu}T^{\nu}+\frac{s}{(k^{2}-s^{2})\boxtimes
(k)}Q^{\mu\nu}\nonumber\\
&  -\frac{s}{(k^{2}-s^{2})\boxtimes(k)}Q^{\nu\mu}+\frac{is^{2}\left(
v_{\alpha}k^{\alpha}\right)  }{k^{2}(k^{2}-s^{2})\boxtimes(k)}\Sigma^{\mu\nu
}+\frac{is^{2}\left(  v_{\alpha}k^{\alpha}\right)  }{k^{2}(k^{2}%
-s^{2})\boxtimes(k)}\Sigma^{\nu\mu}\nonumber\\
&  -\frac{is\left(  v_{\alpha}k^{\alpha}\right)  }{k^{2}(k^{2}-s^{2}%
)\boxtimes(k)}\Phi^{\mu\nu}+\frac{is\left(  v_{\alpha}k^{\alpha}\right)
}{k^{2}(k^{2}-s^{2})\boxtimes(k)}\Phi^{\nu\mu}\biggr\}, \label{Prop_A}%
\end{align}%
\begin{equation}
\text{ }\langle\varphi\varphi\rangle\text{ }=\frac{i}{\boxtimes(k)}\left[
k^{2}-s^{2}\right]  , \label{Prop_phi}%
\end{equation}%
\begin{equation}
\langle\varphi A^{\alpha}\left(  k\right)  \rangle=-\frac{i}{\boxtimes
(k)}\left[  -T^{\alpha}+sv^{\alpha}-\frac{s\left(  v_{\mu}k^{\mu}\right)
}{k^{2}}k^{\alpha}\right]  , \label{Prop_Aphi}%
\end{equation}%
\begin{equation}
\langle A^{\alpha}\left(  k\right)  \varphi\rangle=-\frac{i}{\boxtimes
(k)}\left[  T^{\alpha}+sv^{\alpha}-\frac{s\left(  v_{\mu}k^{\mu}\right)
}{k^{2}}k^{\alpha}\right]  ,
\end{equation}
where: $T_{\mu}=S_{\mu\nu}v^{\mu},$ $S_{\mu\nu}=\varepsilon_{\mu\kappa\nu
}\partial^{\kappa}$, $\ \theta_{\mu\nu}=g_{\mu\nu}-\omega_{\mu\nu},$
$\ \omega_{\mu\nu}=\partial_{\mu}\partial_{\nu}/\square.$

The term, $\boxtimes(k)=\left[  k^{4}-\left(  s^{2}-v.v\right)  k^{2}-\left(
v.k\right)  ^{2}\right]  ,$ determines the pole structure associated with the
poles of such propagators. In ref. \cite{Manojr}, a consistency analysis
investigating the causality, stability and unitarity of such a model was also performed.

\section{Classical Wave Equations and Solutions}

Let us now consider the reduced model, given by Lagrangian (\ref{Lagrange2}),
without the gauge-fixing term:
\begin{equation}
\mathcal{L}_{1+2}=-\frac{1}{4}F_{\mu\nu}F^{\mu\nu}+\frac{1}{2}\partial_{\mu
}\varphi\partial^{\mu}\varphi-\frac{s}{2}\epsilon_{\mu\nu k}A^{\mu}%
\partial^{\nu}A^{k}+\varphi\varepsilon_{\mu\nu k}v^{\mu}\partial^{\nu}%
A^{k}+A_{\mu}J^{\mu}+\varphi J, \label{Reduced}%
\end{equation}
where one observes the Chern-Simons term (having $s$ as topological mass) and
the Lorentz-violating term that couples the fixed 3-vector $v^{\mu}$ to the
gauge vector $A^{\mu}.$ Associated to this Lagrangian there are two
Euler-Lagrangian equations of motion:%

\begin{align}
\partial_{\nu}F^{\mu\nu}  &  =-\frac{s}{2}\varepsilon^{\mu\nu\rho}%
\partial_{\nu}A_{\rho}-\varepsilon^{\mu\nu\rho}v_{\nu}\partial_{\rho}%
\varphi-J^{\mu},\label{motion1}\\
\square\varphi &  =\epsilon_{\mu\nu k}v^{\mu}\partial^{\nu}A^{k}+J.
\label{motion2}%
\end{align}

The modified Maxwell equations associated with this Lagrangian read as below:
\begin{align}
\overrightarrow{\nabla}\times\overrightarrow{E}+\partial_{t}B\text{ }  &
=0,\label{Maxwell1}\\
\partial_{t}\overrightarrow{E}-\nabla^{\ast}B\text{ \ }  &  =-\overrightarrow
{j}+s\overrightarrow{E}^{\ast}+\left(  \overrightarrow{v}^{\ast}\partial
_{t}\varphi+v_{0}\overrightarrow{\nabla}^{\ast}\varphi\right)
,\label{Maxwell2}\\
\overrightarrow{\nabla}.\overrightarrow{E}\text{ }+\text{ }sB\text{ }  &
=\text{ }\rho-\overrightarrow{v}\times\overrightarrow{\nabla}\varphi
,\label{Maxwell3}\\
\square\varphi-\overrightarrow{v}\times\overrightarrow{E}  &  =-v_{0}%
\overrightarrow{\nabla}\times\overrightarrow{A}+J, \label{Maxwell4}%
\end{align}
where the first equation stems from the Bianchi identity\footnote{In $D=1+2$
the dual tensor, defined as $F^{\mu\ast}=\frac{1}{2}\epsilon^{\mu\nu\alpha
}F_{\nu\alpha},$ is a 3-vector given by: $F^{\mu\ast}=(B,-\overrightarrow
{E}^{\ast}).$ Here one adoptes the following convection: $\epsilon
_{012}=\epsilon^{012}=\epsilon_{12}=\epsilon^{12}=1. $ The symbol $(^{\ast}),$
in a general way, also designates the dual of a 2-vector: $\left(
E^{i}\right)  ^{\ast}=\epsilon_{ij}E^{j}\longrightarrow\overrightarrow
{E}^{\ast}=(E_{y},-E_{x}).$} $(\partial_{\mu}F^{\mu\ast}=0)$,$\ $while the two
inhomogeneous ones come from the motion equation (\ref{motion1}), and the last
one is derived from eq. (\ref{motion2}). Explicitly, one notes that
Eq.(\ref{motion2}) can be written as two simpler equations whether the vector
$v^{\mu}$ is purely space- or time-like: $\square\varphi=\overrightarrow
{v}\times\overrightarrow{E}+J,$ \ for \ $v^{\mu}=(0,\overrightarrow{v});$
$\square\varphi=-v_{0}\overrightarrow{\nabla}\times\overrightarrow{A}+J,$
\ for \ $v^{\mu}=(v_{0},\overrightarrow{0})$. Applying the differential
operator, $\partial_{\mu},$ on the eq. (\ref{motion1}), there results the
following equation for the gauge current: $\partial_{\mu}J^{\mu}%
=-\varepsilon^{\mu\nu\rho}\partial_{\mu}v_{\nu}\partial_{\rho}\varphi$, which
reduces to the conventional current-conservation law, $\partial_{\mu}J^{\mu
}=0$, when $v^{\mu}$ is constant or has a null rotational $(\varepsilon
^{\mu\nu\rho}\partial_{\mu}v_{\nu}=0)$. These conditions correspond exactly to
the ones that lead to a gauge invariant theory \cite{Jackiw}.

Manipulating the Maxwell equations, one notes that the fields $B$,
$\overrightarrow{E,}$ satisfy inhomogeneous wave equations:
\begin{align}
(\square+s^{2})B  &  =s\rho+\overrightarrow{\nabla}\times\overrightarrow
{j}-s\overrightarrow{v}\times\nabla\varphi-\partial_{t}\left(  \nabla
\varphi\right)  \times\overrightarrow{v}^{\ast}+v_{0}\nabla^{2}\varphi
,\label{B1}\\
(\square+s^{2})\overrightarrow{E}  &  =-\overrightarrow{\nabla}\rho
-\partial_{t}\overrightarrow{j}-s\overrightarrow{j}^{\ast}-s\overrightarrow
{v}\left(  \partial_{t}\varphi\right)  -sv_{o}\overrightarrow{\nabla}%
\varphi+\overrightarrow{v}^{\ast}\partial_{t}^{2}\varphi+v_{0}\overrightarrow
{\nabla}^{\ast}\left(  \partial_{t}\varphi\right) \nonumber\\
&  +\overrightarrow{\nabla}(\overrightarrow{v}\times\overrightarrow
{\nabla\varphi}), \label{E1}%
\end{align}
which, in the stationary regime, are reduced to:
\begin{align}
(\nabla^{2}-s^{2})B  &  =-s\rho-\overrightarrow{\nabla}\times\overrightarrow
{j}+s\overrightarrow{v}\times\nabla\varphi-v_{0}\nabla^{2}\varphi
,\label{Bfield}\\
(\nabla^{2}-s^{2})\overrightarrow{E}  &  =s\overrightarrow{j}^{\ast
}+\overrightarrow{\nabla}\rho+sv_{o}\overrightarrow{\nabla}\varphi
-\overrightarrow{\nabla}(\overrightarrow{v}\times\overrightarrow{\nabla
\varphi}). \label{Efield}%
\end{align}
Similarly to the behaviour of the\ classical MCS-potential, here the potential
components $(A_{0},\overrightarrow{A})$ obey fourth-order wave equations:%

\begin{align}
\square(\square+s^{2})A_{0}  &  =\square\rho-\square(\overrightarrow{v}%
\times\overrightarrow{\nabla\varphi})-s\overrightarrow{\nabla}\times
\overrightarrow{j}+s\left(  \partial_{t}\overrightarrow{\nabla\varphi}\right)
\times\overrightarrow{v}^{\ast}-sv_{o}\nabla^{2}\varphi,\label{Ascalar}\\
\square(\square+s^{2})\overrightarrow{A}  &  =s\partial_{t}\overrightarrow
{j}^{\ast}+s\overrightarrow{\nabla}^{\ast}\rho+s\overrightarrow{v}\left(
\partial_{t}^{2}\varphi\right)  +sv_{o}\overrightarrow{\nabla}\left(
\partial_{t}\varphi\right)  -s\left(  \overrightarrow{\nabla}(\overrightarrow
{v}\times\overrightarrow{\nabla\varphi})\right)  ^{\ast}\nonumber\\
&  +\square(\overrightarrow{j}-\overrightarrow{v}\partial_{t}\varphi
-v_{0}\overrightarrow{\nabla}^{\ast}\varphi), \label{Avector}%
\end{align}
which are endowed with an inhomogeneous sector much more complex due to the
presence of the terms $\overrightarrow{v}$ and $\varphi$ in the Lagrangian
(\ref{Reduced}). It is instructive to remark that wave equations (\ref{B1},
\ref{E1}, \ref{Ascalar}, \ref{Avector}) reduce to their classical MCS usual
form \cite{Winder},\cite{Ferreira} in the limit one takes $v^{\mu
}\longrightarrow0$, namely:
\begin{align}
(\square+s^{2})B  &  =s\rho+\overrightarrow{\nabla}\times\overrightarrow
{j};\text{ \ }(\square+s^{2})\overrightarrow{E}=-\overrightarrow{\nabla}%
\rho-\partial_{t}\overrightarrow{j}-s\overrightarrow{j}^{\ast};\label{MCS1}\\
\square(\square+s^{2})A_{0}  &  =\square\rho-s\overrightarrow{\nabla}%
\times\overrightarrow{j};\text{ \ }\square(\square+s^{2})\overrightarrow
{A}=s\partial_{t}\overrightarrow{j}^{\ast}+s\overrightarrow{\nabla}^{\ast}%
\rho+\square\overrightarrow{j}. \label{MCS2}%
\end{align}
The above wave equations present the following solutions \cite{Winder} (for a
point-like charge distribution and null current):
\begin{align}
B(r)  &  =\left(  e/2\pi\right)  K_{0}(sr);\text{ \ \ }\overrightarrow
{E}=\left(  e/2\pi\right)  sK_{1}(sr)\overset{\wedge}{r};\label{MCS3}\\
A_{0}(r)  &  =\left(  e/2\pi\right)  K_{0}(sr);\text{ \ \ }\overrightarrow
{A}(r)=\left(  e/2\pi\right)  \left[  1/r-srK_{1}(sr)\right]  \overset{\wedge
}{r^{\ast}}. \label{MCS4}%
\end{align}

Up to now, eq. (\ref{motion2}) was not still used in the derivation of the
wave equations for the fields and potentials. It will be appropriately
considered in the subsequent solutions.

\section{Solutions for the scalar potential and the strength fields in the
static limit}

Wave equation (\ref{Ascalar}), which rules the dynamics of the scalar
potential, $A_{0},$ is already known. This equation, however, is not entirely
written in terms of $A_{0}$, since the scalar field $\varphi$ is not a
constant variable and exhibits its own dynamics described by eq.
(\ref{Maxwell4}), which now must be taken into account to provide the correct
solution of the wave equation. \ Eq. (\ref{Ascalar}) will present two
different solutions depending on the character of the fixed vector $v^{\mu}$,
as we will see below.

\subsection{The external vector is purely time-like: $v^{\mu}=(v_{0},0)$}

Supposing the system reaches a stationary regime, eq. (\ref{Ascalar}) is
reduced to
\begin{equation}
\nabla^{2}(\nabla^{2}-s^{2})A_{0}=-\nabla^{2}\rho-s\overrightarrow{\nabla
}\times\overrightarrow{j}-sv_{o}\nabla^{2}\varphi+\nabla^{2}\left(
\overrightarrow{v}\times\overrightarrow{\nabla}\varphi\right)  .
\label{Ascalar2}%
\end{equation}
In this case, the $\varphi$-field satisfies the equation: $\nabla^{2}%
\varphi=v_{0}B-J.$ \ The use of eq. (\ref{Maxwell4}) changes the eq.
(\ref{Ascalar2}) to the form:
\begin{equation}
\nabla^{2}(\nabla^{2}-s^{2}+v_{0}^{2})A_{0}=-\nabla^{2}\rho-s\overrightarrow
{\nabla}\times\overrightarrow{j}-v_{0}^{2}\rho+sv_{0}J. \label{Azero5a}%
\end{equation}

Starting from a point-like charge-density distribution, $\rho\left(  r\right)
=e\delta(r),$ taking a null current-density, $\overrightarrow{j}=0,$ $J=0,$
and proposing a usual Fourier-transform expression for the scalar potential,
\begin{equation}
A_{0}(r)=\frac{1}{(2\pi)^{2}}\int d^{2}\overrightarrow{k}e^{i\overrightarrow
{k}.\overrightarrow{r}}\widetilde{A}_{0}(k), \label{F}%
\end{equation}
it follows as solution,
\begin{equation}
A_{0}(r)=\frac{e}{(2\pi)w^{2}}\left[  s^{2}K_{0}\left(  wr\right)  +v_{0}%
^{2}\ln r\right]  , \label{Azero3}%
\end{equation}
where: $w^{2}=s^{2}-v_{0}^{2}.$ \ If $\ s^{2}>v_{0}^{2},$ this potential is
always repulsive. Moreover, it is trivial to see that in the limit
$v_{0}\longrightarrow0,$ one recovers the scalar potential associated with the
MCS-Electrodynamics, \ given by eq. (\ref{MCS4}). It is then clear that the
term with dependence on $\ln r$ is then a contribution stemming from the
background field. The electric field, derived from eq. (\ref{Azero3}), is read
as,
\begin{equation}
\overrightarrow{E}(r)=\frac{e}{(2\pi)}\left[  \frac{s^{2}}{w}K_{1}\left(
wr\right)  -\left(  \frac{v_{0}^{2}}{w^{2}}\right)  \frac{1}{r}\right]
\overset{\wedge}{r}, \label{E2}%
\end{equation}
which compared with the MCS correspondent, that of eq. (\ref{MCS3}), possesses
the additional presence of the $1/r$-term, which certainly arises as the
contribution of the background, similarly as it occurs at eq. (\ref{Azero3}).
In the limit of short distance $\left(  r\ll1\right)  ,$ the scalar potential
(\ref{Azero3}), and the electric field (\ref{E2}) are reduced to the form:
\begin{equation}
A_{0}(r)=-\frac{e}{(2\pi)}\left[  \ln r+\frac{s^{2}}{w^{2}}\ln w\right]
,\text{ \ \ \ \ }\overrightarrow{E}(r)=\left(  \frac{e}{2\pi}\right)  \frac
{1}{r}\overset{\wedge}{r},
\end{equation}
which reveals the repulsive character of expression (\ref{Azero3})\ and a
radial $1/r$ electric field near the origin. At the same time, one notices
that, at the origin, the correction-terms induced in eqs. (\ref{Azero3},
\ref{E2}) by the background exhibit the same functional behaviour as the
pre-existent MCS\ terms. However, when one goes far away from the origin, the
picture dramatically changes: the correction-terms entirely dominate over the
exponential-decaying Bessel functions, resulting in the following forms:
\[
A_{0}(r)=\left[  \frac{ev_{0}^{2}}{(2\pi)w^{2}}\right]  \ln r,\text{
\ \ \ }\overrightarrow{E}(r)=-\left[  \frac{e}{(2\pi)}\frac{v_{0}^{2}}{w^{2}%
}\right]  \frac{1}{r}\overset{\wedge}{r},
\]
In this way, one has a substantial change in the asymptotic behaviour of
the\ solutions, so that it is now manifest that one of the main roles of the
background is to promote a sensitive decreasing in the screening (or
decay-factor) of the field solutions. Indeed, a logarithmic scalar potential
and a $1/r$-electric field are usual asymptotic solutions in a massless
QED$_{3}$.

In the absence of currents, the magnetic field is ruled by eq. (\ref{Bfield}),
which reads simply as: $(\nabla^{2}-s^{2}+v_{0}^{2})B=-s\rho.$ This
differential equation is fulfilled by a very simple solution:
\begin{equation}
B(r)=\left(  \frac{es}{2\pi}\right)  K_{0}(wr). \label{B4}%
\end{equation}
In comparing this magnetic field with that of eq. (\ref{MCS3}), one does not
observe any additional term. In this case, the influence of the background
seems to be totally absorbed into the decay factor, $w,$ here smoothly
diminished by the effect of the background. In this form, one remarks that
decisive effects coming from the background are confined to the
electric-sector of the theory. Finally, one can remark that the results here
obtained do not exhibit any signal of spatial anisotropy, which is consistent
with the adoption of a null vector $\overrightarrow{v}$, since this vector is
the element responsible by the choice of a privileged direction in space. The
anisotropy, therefore, must be manifest when $v^{\mu}$ is space-like.

\subsection{The external vector is\ purely\ space-like: $v^{\mu}=(0,v)$}

In this case, the equation fulfilled by the scalar field, $\nabla^{2}%
\varphi=-\overrightarrow{v}\times\overrightarrow{E},$ can be read in term of
the scalar potential: $\nabla^{2}\varphi=\overrightarrow{v}\times
\overrightarrow{\nabla}A_{0}-J=(\overrightarrow{v}.\overrightarrow{\nabla
}^{\ast})A_{0}-J$. Taking into account this relation, eq. (\ref{Ascalar}) in
its stationary regime is reformulated as:
\begin{equation}
\left[  \nabla^{2}(\nabla^{2}-s^{2})-(\overrightarrow{v}.\overrightarrow
{\nabla}^{\ast})(\overrightarrow{v}.\overrightarrow{\nabla}^{\ast})\right]
A_{0}=-\nabla^{2}\rho-s\overrightarrow{\nabla}\times\overrightarrow
{j}-(\overrightarrow{v}.\overrightarrow{\nabla}^{\ast})J, \label{Azero6a}%
\end{equation}
where it was used the relation: $\nabla^{2}(\overrightarrow{v}\times
\overrightarrow{\nabla\varphi})=(\overrightarrow{v}.\overrightarrow{\nabla
}^{\ast})\nabla^{2}\varphi=(\overrightarrow{v}.\overrightarrow{\nabla}^{\ast
})(\overrightarrow{v}.\overrightarrow{\nabla}^{\ast})A_{0},$ since
$\overrightarrow{v}=cte.$

Starting from a point-like charge density distribution, $\rho\left(  r\right)
=e\delta(r),$ $\overrightarrow{j}=J=0,$ and proposing the same
Fourier-transform expression, given by eq. (\ref{F}), one writes:%

\begin{equation}
A_{0}(r)=-\frac{e}{\left(  2\pi\right)  ^{2}}\int_{0}^{\infty}kdk\int
_{0}^{2\pi}d\varphi\frac{e^{ikr\cos\varphi}}{\left[  (\overrightarrow{k}%
^{2}+s^{2})+\overrightarrow{v}^{2}\sin^{2}\alpha\right]  },
\end{equation}
where $\alpha$ is the angle defined by: $\overrightarrow{v}.\overrightarrow
{k}=vk\cos\alpha.$ An exact result was not found for this full integral, but
an approximation can be accomplished in order to solve it algebraically.
Indeed, considering $s^{2}\gg v^{2}$ an integration results feasible. Here,
there is an external vector, $\overrightarrow{v},$ that fixes a direction in
space and the coordinate position, $\overrightarrow{r},$ where one measures
the fields. One then considers that the angle between $\overrightarrow{v}$ and
$\overrightarrow{r}$ is given by: $\overrightarrow{v}.\overrightarrow
{r}=vr\cos\beta,$ where $\beta=cte.$ Considering this information and working
in limit in which $s^{2}\gg v^{2}$, the integration becomes feasible, so that
one attains (at first order on $v^{2}/s^{2}$): \
\begin{equation}
A_{0}(r)=\frac{e}{(2\pi)}\left[  K_{0}(sr)-\frac{\left(  1-\cos^{2}%
\beta\right)  }{2s}v^{2}rK_{1}(sr)+\frac{v^{2}}{2s^{2}}\left(  1-2\cos
^{2}\beta\right)  K_{2}(sr)\right]  . \label{Abeta}%
\end{equation}
In this expression, one notes a clear dependence of the potential on the angle
$\beta,$ which is a unequivocal sign of anisotropy determined by the ubiquity
of background vector on the system. Near the origin, the $K_{2}$-function
dominates over the other terms, so that the short-distance potential behaves
effectively as:%

\begin{equation}
A_{0}(r)=\frac{e}{(2\pi)}\left[  \left(  1-\cos^{2}\beta\right)  \frac{v^{2}%
}{s^{2}}\frac{1}{r^{2}}\right]  ,
\end{equation}
which shows that the potential is always repulsive at origin. In spite of this
fact, the expression (\ref{Abeta}) may exhibit an attractive well region, at
larger $r$-values, depending on the value of the $s$ parameter. This fact
brings into light the possibility of occurrence of pair-condensation
concerning two particles interacting by means this gauge field. This issue
should be more properly investigated in the context of the a low-energy
two-particle scattering \cite{Manojr2}, whose amplitude can be converted into
the interaction potential by a Fourier transform.

Looking at the expression (\ref{Avector}) for the vector potential, one
observes the presence of $\overrightarrow{\nabla}(\overrightarrow{v}%
\times\overrightarrow{\nabla\varphi})$, which can not be written as a term
depending directly on $\overrightarrow{A}$. This fact seems to prevent a
solution for $\overrightarrow{A}$ starting from the static version of this
differential equation, which also seems to be an impossibility for determining
a solution for the magnetic field. However, one must be indeed interested in
the magnetic field, and a simpler solution for it can arise from the eq.
(\ref{Maxwell2}), which in the static regime is simplified to the form:
$\nabla B$ \ $=-s\overrightarrow{E}-v_{0}\overrightarrow{\nabla}\varphi$. For
a pure space-like $v^{\mu}$, this last equation reduces to: $\nabla
B=-s\overrightarrow{E}=s\nabla A_{0}$, an equation that links the magnetic
field to the scalar potential: $B=sA_{0}+cte.$ Based on eq. (\ref{Abeta}), we
achieve the following expression for the fields:%

\begin{align}
\overrightarrow{E}(r)  &  =\frac{e}{(2\pi)}\biggl\{sK_{1}(sr)+\left(
1-\cos^{2}\beta\right)  \frac{v^{2}}{2}\left[  r-\frac{2}{s^{2}r}\right]
K_{0}(sr)\nonumber\\
&  +(1-\cos^{2}\beta)\frac{v^{2}}{2s}\left[  1-\frac{4}{s^{2}r^{2}}\right]
K_{1}(sr)\overset{\wedge}{r}\biggr\},\label{Esplike}\\
B(r)  &  =\frac{e}{(2\pi)}\left[  sK_{0}(sr)-\left(  1-\cos^{2}\beta\right)
\frac{v^{2}}{2}rK_{1}(sr)+\frac{v^{2}}{2s}\left(  1-2\cos^{2}\beta\right)
K_{2}(sr)\right]  . \label{Bsplike}%
\end{align}
Here, the effect of the background vector, $\overrightarrow{v},$ appears more
clearly on the field solutions. As compared to the MCS fields ($B$ and
$\overrightarrow{E}),$ there arise supplementary terms, proportional to
$\cos^{2}\beta,$ responsible by the spatial anisotropy.

\section{\ Final Remarks}

Starting from a dimensionally reduced gauge invariant, Lorentz and
CPT-violating planar model, derived from the Carroll-Field-Jackiw term
(defined in $1+3$ dim.), we have studied the extended Maxwell equations (and
the corresponding wave equations for the field-strengths and potentials)
stemming from the planar Lagrangian. While the field-strengths satisfy
second-order inhomogeneous wave equations, the potential components
($A_{0},\overrightarrow{A})$ fulfill fourth-order wave equations, a clear
similarity to the usual behaviour inherent to the pure MCS sector. As
expected, this structural resemblance is also manifest in the solutions to
these equations. Indeed, in the case of a purely time-like background, one has
attained solutions for the fields $B$ and $\overrightarrow{E}$ that differ
from the MCS counterparts just by correction-terms (dependent on $v^{\mu}$).
These new terms do not bring about any appreciable physical change near the
origin, where they present the same behaviour as the MCS terms. Away from the
origin, however, the panorama is new and intriguing: the correction-terms,
independently of the value of their coefficients, come to dominate over the
MCS behaviour, yielding an appreciable physical change on the asymptotic
solutions. In fact, at large distances, the electric sector of the massive
MCS-Electrodynamics, characterized by strong screening (stemming from the
topological mass), is smoothly replaced by a physical picture typical of a
long-range massless Electrodynamics. Concerning the magnetic field, no drastic
modification takes place, but the decay-factor suffers a softening (due to the
presence of $v_{0}^{2}$). In this case, no signal of anisotropy was observed,
as expected.

In the pure space-like case, the background field appears more explicitly in
the solutions in the form of spatial anisotropy, a consequence of the
selection of a privileged spatial direction, given by $\overrightarrow{v}$.
The solutions keep the MCS\ reference term, but at the same time present a
complex form, which reflects the anisotropy induced by the background over the
solutions, attained in the approximation $s^{2}>>v^{2}$. The reduction of the
screening for large distances, observed in the time-like case, is here absent.
The scalar potential obtained is always repulsive at the origin, but it may
become attractive at an intermediary well- defined region, depending on the
parameter $s$.

The attractiveness of this potential may be better explored in the realm of a
non-relativistic physical system. In fact, the verification of the consistency
of this model (see ref. \cite{Manojr}) in the case of a space-like background
shows that applications of this study to Condensed-Matter systems is a
reasonable option. In this context, there arises the interesting possibility
of investigating a M\"{o}ller scattering in the low-energy (non-relativistic)
limit. For this task, following a usual procedure in QED$_{3}$ \cite{Belich},
one should include the Dirac sector and consider suitable couplings
(electron-photon and electron-scalar ones). The electron-electron interaction
would then be mediated by the scalar and gauge fields, whose action must
appear in the form of an interaction-potential (stemming from a tree-level
calculation) \cite{Manojr2}.

\section{Acknowledgments}

Manoel M. Ferreira Jr is grateful to the Centro Brasileiro de Pesquisas
F\'{i}sicas - CBPF, for the kind hospitality. \ J. A. Helay\"{e}l-Neto
acknowledges the CNPq for financial support.

\end{document}